# Microstructure, Mechanical Properties and Aging Behaviour of Nanocrystalline Copper – Beryllium alloy

Ivan Lomakin[1*], Miguel Castillo-Rodríguez[2] and Xavier Sauvage[3]

[1] *Saint Petersburg University, 199034 Saint Petersburg, Russia*

[2] *IMDEA Materials Institute, 28906 Getafe, Madrid, Spain*

[3] *Normandie Univ, UNIROUEN, INSA Rouen, CNRS, Groupe de Physique des Matériaux, 76000 Rouen, France*

*Corresponding Author: e-mail: ivan.v.lomakin@gmail.com, address: Universiyeysky pr. 28, 198504 Saint Petersburg, Russia

**Abstract**

A complex study of aging kinetics for both coarse-grained and nanostructured by severe plastic deformation Cu – 2 wt.% Be alloy is reported. It is shown that aging of a coarse-grained alloy leads to continuous formation of nanosized CuBe body centred cubic (bcc, CsCl – type) semi-coherent particles with the {220} Cu // {200} CuBe crystallographic orientation relationship. These particles created significant internal stress fields and became obstacles for dislocation glide that resulted in a change in the hardness from 95 Vickers hardness (HV) for the solubilized alloy to 400 HV for the aged one. The severe plastic deformation led to the formation of a single-phase nanograined microstructure with an average grain size of 20 nm and 390 HV. It was found that this grain size was slightly driven by grain boundary segregation. Further aging of the nanocrystalline alloy led to the discontinuous formation of precipitates on the former Cu grain boundaries and skipping of metastable phases. Significant age hardening with a maximum hardness of 466 HV for the aged nanostructured alloy was observed. Mechanical tests result revealed a strong influence of microstructure and further aging on strength capability of the alloy for both coarse-grained and nanostructured alloy. A good thermal stability in the nanostructured alloy was also noticed. Theoretical calculations of the hardness value for the CuBe phase are provided. It was shown that Be as a light alloying elements could be used for direct change of microstructure ~~and aging behaviour~~ of severely deformed copper alloys.

# 1. Introduction

Nanostructured copper alloys often demonstrate the interesting combination of electrical conductivity and mechanical strength that makes them good candidates for future engineering applications [1]. Nanostructuring of copper alloys may also improve wear resistance [2,3], radiation resistance [4] and magnetic properties [5,6]. A popular way to achieve ultrafine grained (UFG) or nanoscale structures (NS) is the application of severe plastic deformation (SPD) processes (such as equal channel angular pressing (ECAP) or high-pressure torsion (HPT)) that have been intensively studied during the last two decades [7]. One should note that such processes can also lead to specific strain induced structural features, such as grain boundary segregation [8], phase transitions [6,9,10] or dynamic precipitation [11,12]. Moreover, the precipitation sequence and the aging kinetics are often significantly affected in such UFG structures [11,13–18]. The influence of SPD on grain refinement and precipitation has been reported for Cu – Cr, Cu – Co and Cu – Ag alloys [2,3,5,6,10]. However, most of the studies of nanostructured binary Cu alloys rely on X-ray diffraction (XRD) and little is reported about the exact mechanisms of phase separation and precipitation.

The Cu – Be system is another interesting candidate [19] because such alloys in the classical coarse-grain (CG) state provide an attractive compromise between functional and mechanical properties, such as corrosion resistance, hardness, tensile strength, fatigue endurance and electric conductivity [20]. Also, there is a large size mismatch between the Be and Cu atoms that may lead to strong interactions between the beryllium atoms and crystalline defects, such as dislocations and grain boundaries (GB). This should significantly affect the precipitation mechanisms in the UFG state and the resulting properties. The properties of Cu – Be alloys are indeed closely linked to the precipitation of Be-rich nanoscale particles through aging [21]. The classical treatment involves a solution treatment in a temperature range of 700°C to 800°C followed by quenching and then a low-temperature aging in a range of 280°C to 350°C. As a result, a material with a hardness up to 400 HV, which corresponds to a 1100 MPa yield stress, and electrical conductivity of 15% - 20% IACS can be achieved [21]. The following precipitation sequence in coarse-grained Cu – Be alloys has been established:

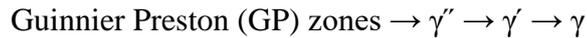

Guinnier Preston (GP) zones → γ″ → γ′ → γ

where the γ phase is the stable CuBe phase (CsCl – type) whereas the GP zones, bcc tetragonal monoclinic γ" and the bcc γ' are metastable phases [22,23]. This precipitation sequence can be significantly affected by other alloying elements. Thus addition of Co as an alloying element may prevent nucleation of γ phase [24] while addition of Zn leads to skipping metastable γ" phase [25]. On the other hand, aging behavior of Cu – Be alloys also dependent on the exact aging conditions. For example, an increase in the heating rate promotes the direct nucleation of the stable γ phase. A complex differential scanning calorimetry study [24] of Cu – 2 wt.% Be alloy showed than low and medium heating rates (up to 30°C/min) lead to simultaneous discontinuous and continuous precipitation processes, while an increase of the heating rate promotes the direct continuous precipitation of the stable γ phase. The complicated influence of continuous precipitation on discontinuous precipitation was revealed in [26] and seems to be controlled by the cell volume fraction driven by aging temperature ranges.

The aim of the present study was to investigate how it could be possible to combine precipitate hardening and GB strengthening in the Cu – Be system that results from nanoscale structures that are obtained by a combination of SPD and thermal treatments. A commercial Cu – 2Be alloy was first solutionised to investigate the influence of Be in solid solution on the SPD induced grain refinement mechanisms. Then, special emphasis was given to the influence of the ultrafine grain size on the precipitation kinetics and mechanisms.

## 2. Materials and Methods

A hot rolled Cu – 2 wt. % Be rod (diameter of 20 mm) was annealed at 780°C for 30 minutes and water quenched to obtain a Be super-saturated solid solution. Annealed discs (thickness 2 mm, diameter 20 mm) were processed by HPT under a pressure of 6 GPa and up to 5 revolutions at 1 rpm using a Walter Klement GmbH HPT-008 device. Aging treatments in the temperature range of 225°C to 375°C were carried out from 3 minutes up to 10 hours in a Nabertherm N40E muffle furnace. Specimens were placed in an alumina crucible filled with silicon dioxide sand during annealing. An external thermocouple with accuracy of 0.1°C was used to control the temperature of the sand. Microhardness measurements were carried out with a Shimadzu HVG – 21 microhardness testing machine using a Vickers indenter with a load of 0.5 kg and a dwell time of 15 seconds. The HV was measured at the distance of 5 mm from the centre of HPT discs. The samples were ground and polished with a 50 nm $SiO_2$ final suspension. Each microhardness value is an average of ten indentations. An Olympus BX51 optical microscope was used to determine the grain morphology and size distribution over 300 grains in the CG material.

Differential scanning calorimetry (DSC) analyses were performed using a NETZSCH DSC 204 F1 Phoenix calorimeter to study thermal processes that occurred in the CG and NS states. Specimens of 40 mg were subjected to two thermal cycles in the temperature range of 20°C to 550°C with a heating/cooling rate of 10°C/min and empty aluminium crucibles were used for reference. After the first cycle, a second was made with the same conditions such that the DSC curves provided in the present work were obtained by subtracting the two curves obtained during the first and second cycles.

XRD studies were carried out using a Bruker D8 Discover high-definition X-Ray diffractometer with CuKα radiation. The 2θ Bragg angle was varied from 25º to 125º with a scan rate of 5º/min. The XRD measurements for the NS alloys were carried out on the cross section of the HPT discs at a distance of 5 mm from the centre. Prior to the measurements, samples were ground, polished and etched with a solution of 100 ml distilled water and 10 mg ammonium persulfate for 30 seconds.

Transmission Electron Microscopy (TEM) specimens were prepared from 3 mm discs cut 5 mm from the HPT disc centre. Specimens were then mechanically ground to a thickness of ~100 μm and a Struers TenuPol – 5 twin jet electrolytic polisher was finally used to reach electron transparency. The electrolyte consisted of 1/3 $HNO_3$ + 2/3 methanol at a temperature -30ºC. The TEM observations were carried out on a FEG S/TEM microscope (Talos F200X, FEI) with a point resolution ≤ 0.25 nm, information threshold ≤ 0.12 nm and a STEM-HAADF resolution ≤ 0.16 nm at 200 kV. Additional data were obtained by Scanning TEM (STEM) using a JEOL ARM-200F operated at 200kV. Dark Field (DF – collection angles 20 to 80 mrad) and high-angle annular dark field (HAADF – collection angles 80 to 180 mrad) images were recorded with a probe size of 0.2 nm and a convergence angle of 34 mrad.

Measurements of the mechanical properties of the alloy subjected to different thermomechanical treatments were carried out via uniaxial tensile tests over sub-sized specimens. An experimental technique described in [27] using a Shimadzu AGX-50 Plus universal mechanical testing machine with strain rate of $5 \times 10^{-4}$ $s^{-1}$ was used. The mechanical parameters presented in the paper are averaged over three tests.

## 3. Results and discussion

### 3.1 Precipitation in the coarse-grain state

After the solution treatment, the Cu – 2Be alloy exhibited a coarse-grained structure with a mean grain size of 25 μm and few recrystallization twins (data not shown here). After artificial aging for 10 h at 325°C, a high density of dark particles is clearly exhibited in the HAADF images (Fig. 1a). This contrast indicates that they contain a large Be concentration (Z-contrast), as expected [22,23]. The particles are elongated with a lens shape with a mean length of about 30 nm and mean thickness of about 5 nm. They are oriented in various specific directions, indicating that they have a specific orientation relationship with the matrix with several variants. The high resolution HAADF STEM image (Fig. 1b) and related fast Fourier Transforms (FFT) show that they could be attributed to semi-coherent CuBe γ phase particles (bcc lattice, CsCl-type) with {220}Cu // {200} CuBe. This is consistent with the classical orientation relationship for a bcc particle in a fcc matrix [28] and also with data previously published by other authors about the Cu – Be system [22,23]. The precipitates are elongated toward the <110> Cu direction, which does not correspond to the <001> elastically soft direction [29]. Thus, this growing direction is most probably favoured by the relatively small misfit (about 6%) between the {220} Cu (1.35Å) and {200} CuBe (1.27Å) atomic planes. This growth direction creates significant lattice strains in the matrix, and in combination with precipitate hardening, leads to the typical high strength reported for such alloys in peak aged conditions (1100 MPa yield stress and 390 HV [21]).

### 3.2 Nanostructure achieved by HPT

The microstructure of the Cu – 2Be alloy after HPT was relatively homogeneous and characterized by a nanoscale grain size (Fig. 2a and Fig. 2b) The grain size distribution was quite narrow and ranged from 10 to 30 nm with a mean value of about 20 nm (Fig. 2c). It is known that grain refinement during SPD of fcc metals such as copper and aluminium is very sensitive to the level of impurity. Thus, high purity copper processed by SPD typically exhibits an average grain size in the range of 380 – 470 nm which depends on preliminary thermal treatment condition [30], while a typical value of about 150 nm is reported for coarse-grained copper with 1.49 wt.% Si deformed in similar conditions [31]. The Si impurity remain in the solution with Cu at this composition [32]. Thus, impurities dissolved in copper matrix prior to SPD introduce a pinning effect that inhibits dynamic recovery and leads to finer microstructure. One can conclude that Be atoms in solid solution strongly affected the dynamic recovery processes of the defects leading to a smaller grain size. On the diffraction pattern (Fig. 2a, inset), only Debbye-Scherer rings corresponding to fcc Cu and typical UFG structures are exhibited. This indicates that no dynamic precipitation occurred during deformation and that the Be super saturated solid solution was preserved. However, to reveal the possible Be redistribution during the HPT, the microstructure was also characterized by STEM using BF and a HAADF detectors simultaneously and providing the latter a z-contrast image. Such images are displayed in Fig. 3a and Fig. 3b. Grains near the Bragg conditions are darkly imaged on the BF image (Fig. 3a) and thus, few GBs are clearly shown. ~~On the corresponding HAADF image (Fig. 3b), the contrast is not uniform. Variations over a large length scale of about 100 nm, or more, were attributed to local foil thickness variations resulting from the sample preparation. However.~~ A network of narrow black layers along the GBs is apparent and could be attributed to Be segregation at the GBs. The HAADF line profile across two GBs (Fig. 3c) shows that the typical apparent width of such segregation would be about 2nm. However, considering the projection of the foil thickness combined with the GB roughness and misorientation, this value should be considered as an upper bound.

To determine if the Be content in the fcc Cu lattice was significantly affected by such segregations, the lattice parameter was investigated based on the XRD peak shift. The fcc Cu (111) XRD peaks for the Cu-2Be alloy after different thermomechanical treatments are displayed in Fig. 4. In the CG state, a significant shift toward smaller angles occurred during the aging treatment (corresponding to a lattice expansion, see Table 1) and was directly connected to the Be depletion of the Cu matrix due to the precipitation of the γ phase. One should note that a significant peak broadening is also exhibited in the CG aged material. This is related to the lattice distortions created by the semi-coherent interfaces of the γ phase precipitates, as reported in the previous section. It is, however, interesting to note that the (111) Cu peak of the deformed SPD material is not significantly shifted compared with the CG state. The shift is also negligible for both materials after aging. There is only a significant peak broadening that was attributed to the grain refinement. Thus, most of Be was retained within the solid solution in the UFG structure that was achieved by HPT and only a small fraction segregated along GBs, as revealed by the STEM HAADF images.

If one assumes that GBs are covered by a monolayer of Be atoms, then the fraction of Be atoms at the GBs can be estimated. An estimate of the number of available GB sites is:

$$N_{\text{site}} \sim \frac{2 \cdot S}{a_{\text{Cu}}^2} \qquad (1)$$

where $S$ is the GB surface and $a_{\text{Cu}}$ the lattice parameter of Cu ($a_{\text{Cu}} \sim 0.36$ nm).

Thus, considering that GBs are shared by two neighbouring grains, and that grains are spherical with a diameter $d$, then the number of sites per unit of volume is:

$$N_{\text{site}}^* \sim \frac{6}{(d \cdot a_{\text{Cu}}^2)} \qquad (2)$$

Then, the fraction of Be atoms, $X_{\text{Be}}$, required to cover all GBs with a monolayer is:

$$X_{\text{Be}} \sim N^*_{\text{site}} \cdot V_{\text{at}} \quad (3)$$

where $V_{\text{at}}$ is the mean atomic volume in the fcc lattice, i.e. $V_{\text{at}} = a_{\text{Cu}}^3 / 4$. Thus:

$$X_{\text{Be}} \sim \frac{3 \cdot a_{\text{Cu}}}{2 \cdot d} \quad (4)$$

with a mean grain size of 20 nm, $X_{\text{Be}} \sim 2.7$ at.%. Thus, this estimate shows that the Be super-saturated solid solution of about 12.5 at.% (corresponding to 2 wt.%) is only slightly affected by Be segregation at the GBs and should not give rise to a noticeable XRD peak shift.

### 3.3 Influence of severe plastic deformation and thermal treatment on mechanical properties

Aging treatments at different temperatures (ranging from 225°C to 375°C) were carried out for both CG and NS materials. As expected, they significantly affected the microhardness. For the solution treated CG Cu-2Be alloy (Fig. 5a), the highest hardness value was achieved by annealing at 300°C for 10 hours (402±8 HV against 95±4 HV in the solutionised state). At higher temperature (e.g. 375°C), the peak hardness was achieved in a shorter time (30 min) but is significantly lower (325±9 HV). At a lower temperature, the hardening was delayed due to the lower atomic mobility and slower kinetics. In the NS alloy (Fig. 5b), the hardness reached 390±15 HV prior to aging due to the nanoscale grains. Note that only low temperature aging led to hardening with a maximum value of about 466±3 HV for 30 min at 275°C or 1h at 250°C. In any case, further aging led to a monotonic decrease of the hardness level. Isochronal annealing (for one hour) of the NS state showed that the microhardness value stayed beyond 450 HV up to 275°C, but at higher temperature a dramatically softening occurred. Thus, the aging response of the NS state significantly differs from the CG state and was likely the result of the competition between precipitate hardening and grain coarsening that led to softening.

To shed a light on the influence of aging thermal treatment on mechanical properties for both CG and NS Cu – 2Be alloys, uniaxial tensile tests were carried out. Obtained stress-strain curves are presented in Fig. 6 with extracted mechanical properties data listed in the Table 2. In the coarse-grained super saturated solid solution state after thermal treatment at 780°C for 30 minutes the alloy exhibits a low strength, with a yield stress of 140 MPa and an ultimate tensile strength (UTS) of 440 MPa, accompanied by high ductility (uniform elongation of 75%, Fig. 6a). Such yield limit and UTS values are higher than for pure copper and can be explained by Be in solid solution creating obstacles for dislocation glide. One can see that subsequent artificial aging leads to dramatic changes in mechanical response. Increasing of aging time brings strengthening of the alloy with reduction of capability for plastic deformation. Such tendency is driven by continuous formation of the secondary CuBe phase particles as shown on Fig. 1. Increasing of their size and quantity during aging as well as decreasing of the distance between separated obstacles create effective barriers for dislocation glide. This leads to significant strengthening with a yield stress up to 1170 MPa and a decrease of uniform elongation (down to 4%).

As seen from Fig. 6b a significant strengthening of the Cu – 2Be alloy subjected to severe plastic deformation was observed. The alloy, characterized by single-phased microstructure with average grain size of 20 nm, demonstrates a yield limit of 1050 MPa and 10% elongation to failure. This high ductility capacity is unusual for such small grain size since it is difficult to carry such elongation only by dislocation glide. Thus, one may assume that a significant contribution is brought by grain boundary sliding probably promoted by GB segregation of Be (Fig. 3).

Thus, a significant hardening resulted from the grain size reduction. The contribution from the GB strengthening can be rationalized through the Hall-Petch relationship [33,34]:

$$\Delta\sigma_y = k_y \cdot d^{-1/2} \tag{5}$$

where $\Delta\sigma_y$ is the yield stress increase due to the GB contribution, $d$ the grain size and $k_y$ is a constant. For the case of pure copper ($0.12 < k_y < 0.14$ MPa·m$^{-1/2}$ [35,36]) with a mean grain size of 20 nm (similar to the present work), this law predicts an increase in the yield stress in the range of 850 to 990 MPa. If one adds the contribution of the solid solution (130 MPa, see Table 2), then the estimated theoretical yield stress is in the range of 980 to 1120 MPa which is in good agreement with our experimental data (1050 MPa, see Table 2)

Further aging of SPD-processed alloy for short periods of 3 and 10 minutes made the alloy brittle with a UTS of 1050 MPa and 1250 MPa correspondingly. Annealing at 275°C for a longer period of 10 hours leads to further strengthening up to a yield limit of 1300 MPa and an ultimate tensile stress of 1375 MPa, but still with a poor ductility (elongation of 4%). One can conclude that additional strengthening was brought by precipitation and its effect on the yield limit efficiently balanced the softening that should result from grain growth.

**3.4 Influence of the nanoscale grain size on the aging behaviour and on the precipitation kinetics**

To investigate the influence of the nanoscale grain size on the aging behaviour and on the precipitation kinetics, XRD, DSC and TEM analyses were carried out. The XRD patterns from the CG and NS Cu-2Be alloy at different annealing stages are displayed in Fig. 7. After solution treatment (780°C during 30 min), only peaks related to fcc Cu were detected. Annealing at 325°C led to the progressive nucleation and growth of the bcc CuBe γ phase and fcc Cu peak broadening due to lattice strains, as discussed in the previous section. The NS state shows only broad fcc Cu peaks resulting from the nanoscale grain size. Upon annealing at 275°C, the bcc CuBe γ phase progressively appears, indicating that the final structure contains the same phases as the CG state.

To gain more detail about the precipitation sequence, DSC measurements were carried out. The curves obtained for the CG and NS Cu-2Be alloys are compared in Fig. 8. For the CG case, six peaks are shown and labelled A to F. The onset temperatures corresponding to all of the thermal events are listed in Table 3. They can be separated into two groups, namely the first temperature interval of 100°C to 290°C and the second from 290°C to 480°C. According to previous investigations at the same heating rate [24], these peaks can be identified as follows. Peak "A" is associated with short range atomic motions probably resulting from the fast annihilation of quenched-in vacancies; "B" is related to coherent Guinier-Preston (GP) zones lying in {100} fcc Cu planes; "C" is related to the $\gamma''$ metastable phase with a monoclinic structure; "D" is attributed to discontinuous precipitation leading to the nucleation and growth of the stable γ phase directly from the supersaturated solid solution; "E" is associated with the transformation of GP zones or γ" precipitates into the metastable bcc $\gamma'$ phase and "F" is attributed to the transformation of $\gamma''$ or $\gamma'$ precipitates into the stable bcc γ phase. In the NS Cu – 2Be alloy only three exothermal peaks could be identified, labelled as G, H and I. The peak, I, is located at nearly the same position as D for the CG state, indicating that discontinuous precipitation occurred. However, peaks related to the homogeneous precipitation of metastable phases (B, C and E) are clearly absent. This may be the result of the high density of GBs that may provide nucleation sites for the stable γ phase. Since the NS state exhibits a nanoscale grain size, grain growth that is an exothermic process is expected and probably accounts for the broad peak labelled as H. The analysis of DSC curves of the nanostructured alloy clearly shows that precipitation in severely deformed alloy proceeds by discontinuous mechanism since only peak "I" is exhibited. In Cu-Be alloys discontinuous secondary phase formation proceeds via boundary migration [37], i.e. discontinuous precipitates growth is controlled by grain boundary diffusion of Be [38]. In CG alloys, homogeneously nucleated precipitates are semi-coherent needle-shaped with a typical length of about 40 nm and a typical width of about 5 nm. However, in the nanocrystalline state, moving boundaries (due to grain growth) and fast diffusion of Be atoms along GBs strongly favour discontinuous precipitation. It also leads to a shift of the discontinuous γ-phase formation to lower temperatures in comparison with the CG alloy.

A subsequent short thermal treatment at any temperature led to mobility of crystallographic defects, such as stacking faults and grain boundary dislocations. According to the DSC traces, up to the temperature 275°C (heating for 25 minutes) this process was accompanied by the formation of non-equilibrium preliminary phases, such as GP – like zones and γ" phase on the grain boundaries. Thus, superposition of these two processes, namely defect annihilation and Be atom diffusion, led to the abnormal mechanical behaviour indicated by the HV value discrepancy. Further aging led to incorporated processes of final Cu-Be bcc formation and grain growth.

To clarify the influence of nanostructuring on second phase nucleation and on the aging response, TEM observations were performed. As shown on Fig. 8, after aging at 275°C during 30 min, the mean grain size significantly increased to about 50 nm (Fig. 9a). In addition, the SAED pattern clearly revealed two phases, namely the fcc Cu and bcc γ phase. Bright field (Fig. 9b) and dark field images (Fig. 9c) obtained using a reflection corresponding to the γ phase (arrowed on the SAED set in Fig. 9b) shows that the γ phase particles are relatively more equiaxed than those observed for the CG alloy (see Fig. 1) and nucleated at the GBs. Fig. 8(d) is a HRTEM micrograph of the γ phase particle in Fig. 9c, where the fast Fourier transform FFT (Fig. 9d inset) of the red square is also displayed. The reflection marked corresponds to a d-spacing of ~ 0.26 nm, whose value fits quite well to the {100} d-spacing planes for CuBe. Further aging up to 10 h at 275ºC resulted in a relatively limited grain growth (Figs. 10 and 11a) with a mean grain size of about 70 nm. Stable γ phase particles are still exhibited and the DF image (Figs. 11b and 11c) shows that they have significantly grown as compared to the 10 min aging treatment (Fig. 9). They stand at former grain boundaries of the super-saturated solid solution and have grown to a size similar to the FCC Cu grains (Fig. 11d). Thus, the TEM observations are consistent with DSC measurements. The Cu – 2Be super-saturated solid solution with a nanoscale grain size achieved by SPD exhibited different phase separation mechanisms as compared to the CG state. It occurred at a lower temperature, without the homogeneous nucleation of metastable phases, but directly through the heterogeneous nucleation of the stable γ phase at GBs. It also appears to be concomitant with grain growth.

Surprisingly, even if no intra-granular homogeneous precipitation could be detected, a significant hardening was observed during annealing. The mean grain size significantly grew (390±15 HV for 20 nm in the as-HPT state, 465±4 HV for 50 nm after 10min at 275°C and 413±2 HV for 70 nm after 10h at 275°C).

The equilibrium volume fraction of the γ phase, $f_v^\gamma$, for the Cu-2Be alloys was near 30% and this intermetallic phase contributed to the hardening. As a simple approach which can be used [33,39], the rule of mixtures may be considered, as shown in Equation (6):

$$HV = (1 - f_y^\gamma) \cdot HV_{Cu} + f_y^\gamma \cdot HV_\gamma \qquad (6)$$

where $HV_{Cu}$ is the hardness of the Cu grains that can be estimated as a function of the grain size as shown in Equation (5) and that is also influenced by the solid solution. The $HV_\gamma$ is the hardness of the intermetallic γ phase. Assuming that after 10 min and 10 hours at 275°C, the Be super saturated solid solution is fully decomposed (i.e. no more solid solution hardening in Cu and equilibrium volume fraction of γ phase has been reached), then Equation (6) provides $1045 < HV_\gamma < 1115$ after 10min and $955 < HV_\gamma < 1015$ after 10 hours at 275°C, respectively, in the nanoscale structures. It is much higher than Cu, as expected for an ordered intermetallic phase. It is almost size independent (just a bit harder when particles are smaller after 10 min), which is also consistent with a brittle phase where dislocation activity and pile-ups are very unlikely.

## 4. Conclusions

According to obtained results, following conclusions can be made:

1. A microstructural study via high resolution STEM showed that precipitation in coarse-grained Cu – 2Be alloy attributed to semi-coherent CuBe γ phase nanosized particle formation (bcc lattice, CsCl-type) with the {220}Cu // {002}CuBe crystallographic orientation relationship.

2. The HPT processing of solubilized Cu – 2Be alloy led to the formation of a single-phase nanograined microstructure with an average grain size of 20 nm and 390 HV. The microstructure that was accompanied by the parent copper matrix lattice parameter changes showed that such grain size was only slightly driven by grain boundary segregation.

3. Aging of the as-deformed Cu – 2Be alloy led to secondary discontinuous CuBe particle formation accompanied by grain growth from 20 nm up to 70 nm after aging at 275°C for 10 hours. That result indicates good thermal stability for the nanostructured alloy. A significant hardening up to 450 HV was observed in the nanocrystalline alloy after aging at 275ºC for 30 minutes.

4. The CuBe particle formation kinetics during aging in nanocrystalline alloy was revealed as discontinuous at the lower temperature in comparison with CG alloy. This caused by faster diffusion of Be atoms along GBs of the nano sized grains. The final γ phase bcc stable structure forms at former grain boundaries of the super saturated solid solution skipping metastable phases and have grown to a size similar to fcc Cu grains. A pinning effect from the precipitate particles led to high thermal stability and grain size less than 100 nm even after aging for extended periods.

5. Mechanical properties of the coarse-grained Cu – 2Be alloy are controlled by CuBe secondary phase particles that significantly affect dislocation glide. Severe plastic deformation applied to the solution treated alloy brings a significant strengthening (yield stress up to 1050MPa) attributed to the combination of Hall-Petch and solid solution strengthening. Annealing at 275°C during 10 hours leads to further strengthening (yield stress up to 1300 MPa) attributed to the formation of a nanoscaled dual-phase structure.

**Acknowledgments**

Current study was supported by Saint Petersburg State University via Lot 2017 applied (Id: 26130576). XRD results were obtained by the "Center for X-ray Diffraction Methods" and DSC traces were performed by "Center for Thermogravimetric and Calorimetric Research" of the Research park of Saint Petersburg State University. Anvils for HPT tool were produced in collaboration with center "Applied aerodynamics" of the Research Park of Saint Petersburg State University. I. Lomakin and X. Sauvage are grateful to Embassy of France in Russia for supporting their collaboration through Mechnikov's scholarship.

**Tables**

Table 1. Influence of thermomechanical treatment on Cu (1 1 1) peak position in Cu – 2 wt. % Be alloy with related lattice parameters

|  | CG | Aged CG | HPT | Aged HPT |
|---|---|---|---|---|
| Cu (1 1 1) plane 2θ angle, degrees | 43.82 | 43.45 | 43.85 | 43.46 |
| Cu (1 1 1) d-spacing, Å | 2.0642 | 2.0809 | 2.0629 | 2.0805 |

Table 2. Results of the mechanical tests for coarse grained and severely deformed Cu – 2Be alloy after different aging thermal treatments

|  |  | $\sigma_{0.2}$, MPa | UTS, MPa | δ, % |
|---|---|---|---|---|
| CG |  |  |  |  |
|  | Solid solution | 130 | 450 | 66 |
|  | 325°C for 10 min | 340 | 600 | 35 |
|  | 325°C for 30 min | 780 | 1050 | 15 |
|  | 325°C for 10 hours | 1170 | 1280 | 6 |
| HPT |  |  |  |  |
|  | As processed | 1050 | 1150 | 10 |
|  | 275°C 10 hours | 1300 | 1375 | 4 |

Table 3. Calorimetric exothermal peaks temperatures observed during heating of coarse grained (CG) and nanostructured by high-pressure torsion (NS) Cu – 2 wt.% Be alloy. Alloy was preliminary annealed at 780°C for 1 hour before calorimetric study for CG alloy and before processing for NS alloy. Temperatures are given in Celsium degrees.

|  | A | B | C | D | E | F | G | H | I |
|---|---|---|---|---|---|---|---|---|---|
| CG | 121.3 | 200.8 | 248.3 | 342.5 | 374.3 | 435.5 | - | - | - |
| NS | - | - | - | - | - | - | 148.4 | 249.7 | 328.3 |

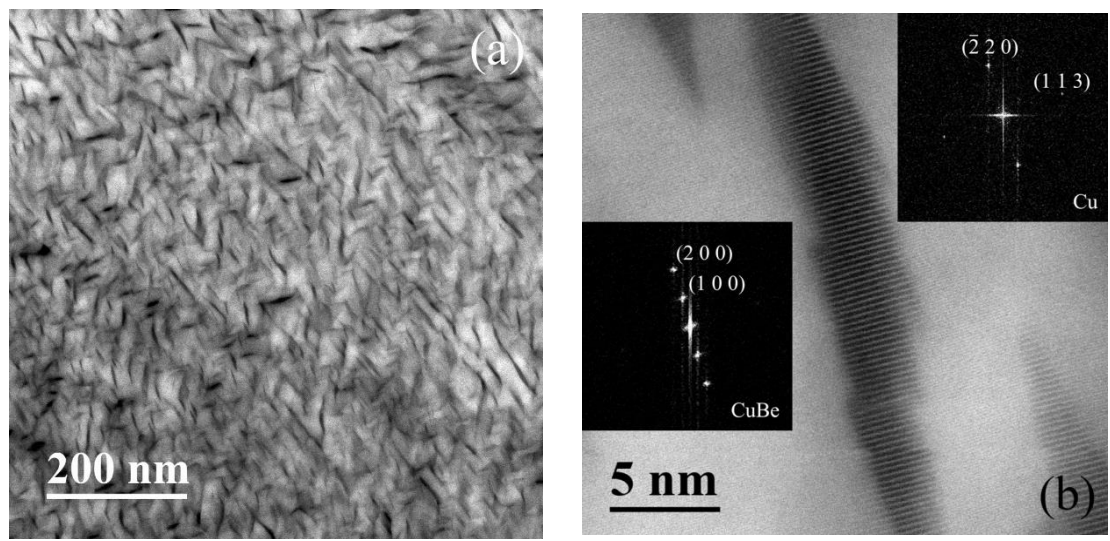

Fig. 1. (a) STEM HAADF image of the CG Cu-2Be alloy aged at 325°C for 10 hours that shows Be-rich lens-shaped particles (dark colour). (b) High resolution STEM HAADF image and corresponding local FFTs for the FCC Cu matrix and in the bcc CuBe phase.

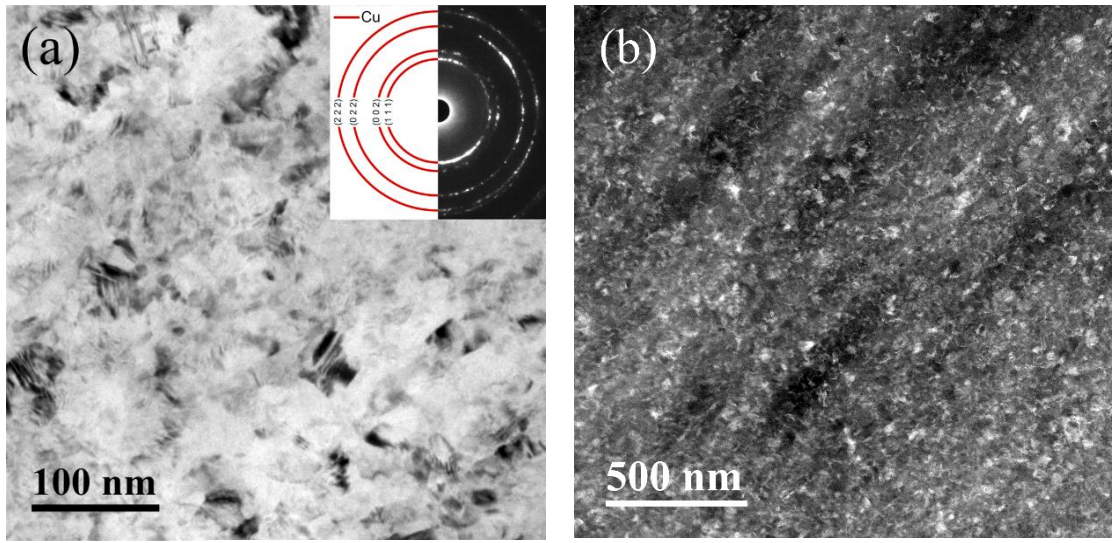

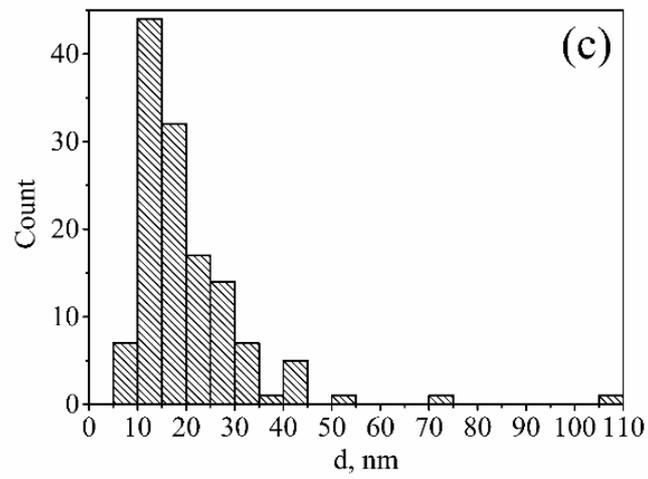

Fig. 2. (a) TEM BF and (b) STEM DF images of the Cu-2 Be alloy, processed by HPT. Inset: SAED obtained with an aperture of 1.2 μm. (c) Grain size distribution histogram in as-HPT processed Cu-2Be alloy.

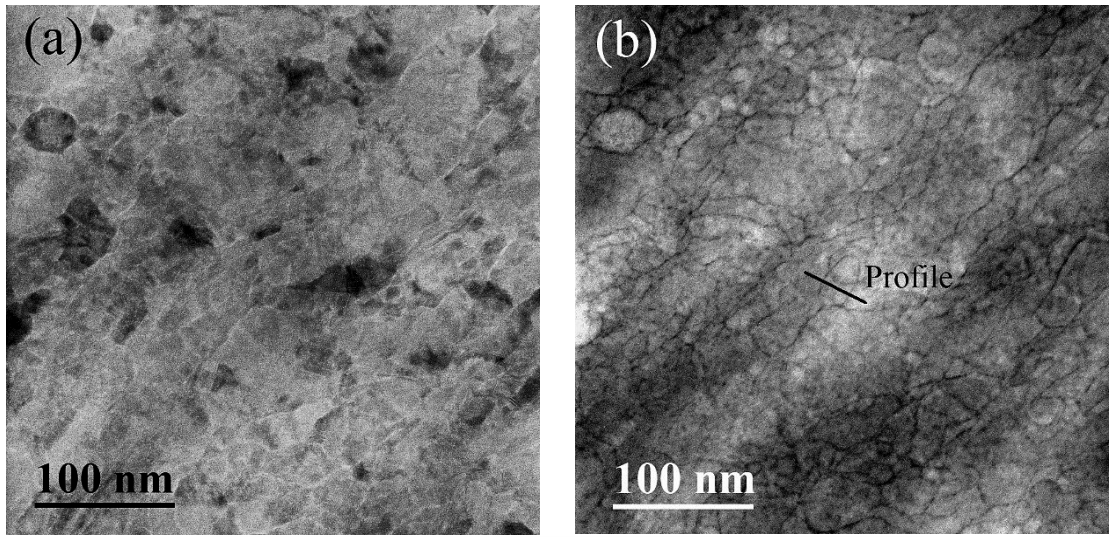

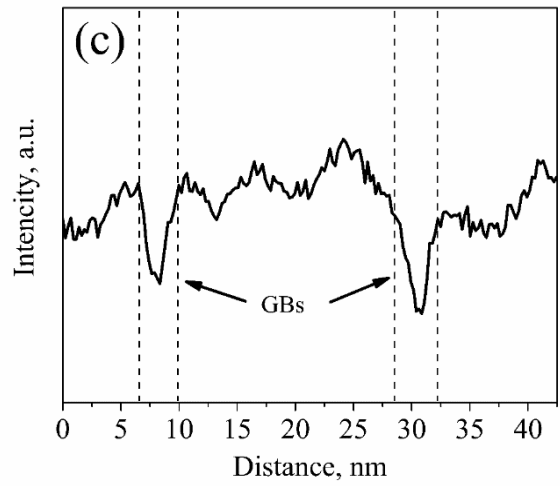

Fig. 3. (a) STEM BF and (b) STEM HAADF images of the Cu-2Be alloy, subjected to HPT. (c) Profile contrast marked with line on Fig. 2(b).

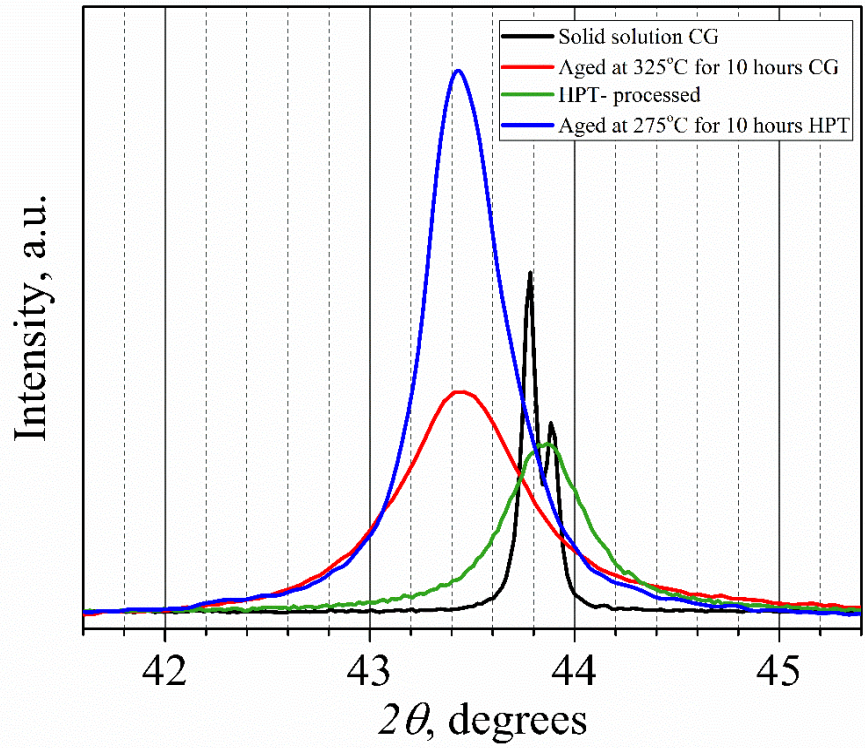

Fig. 4. The influence of thermomechanical treatment on (111) copper matrix peak in the Cu-2 wt. % Be XRD patterns.

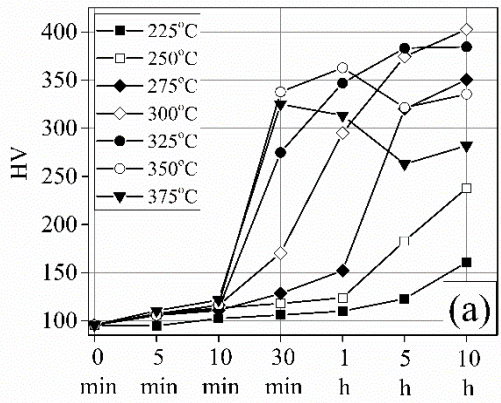

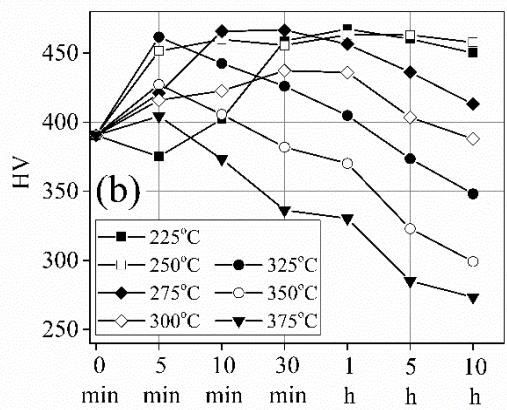

Fig. 5. Artificial aging kinetics for the (a) coarse-grained and (b) nanostructured Cu-2Be alloy by high pressure torsion (HPT).

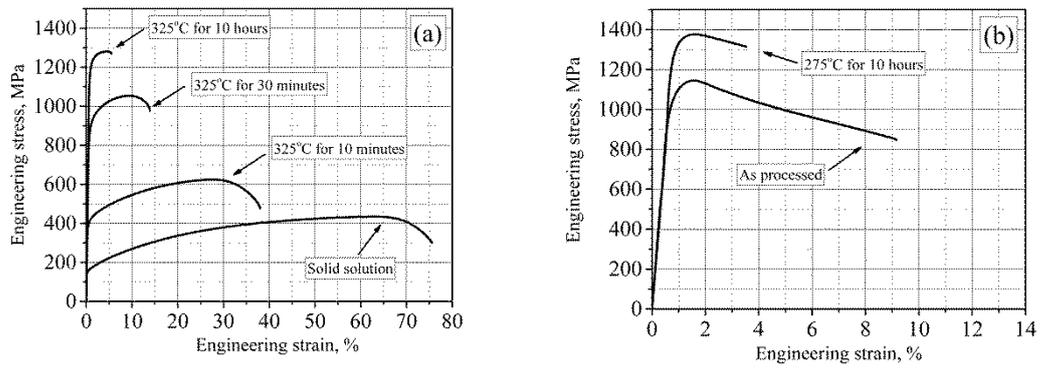

Fig. 6. Engineering stress-strain tensile curves for the coarse-grained (a) and nanostructured (b) Cu – 2Be alloy after artificial aging for the different time at the temperatures 325°C and 275°C correspondingly.

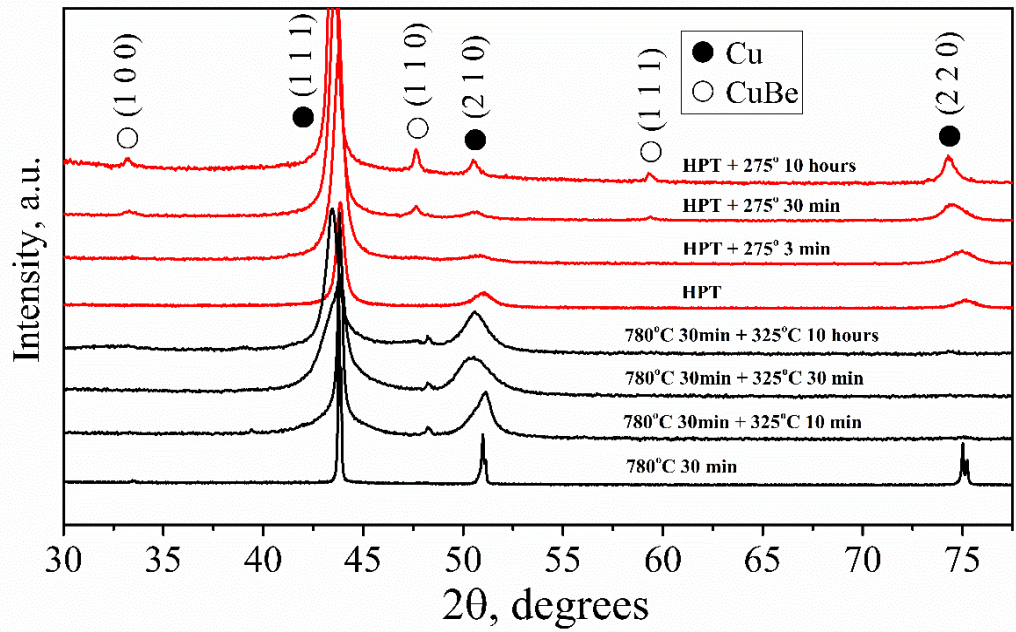

Fig. 7. XRD patterns of CG (black curves) and NS formed by HPT (red curves) for Cu-2Be alloy after different thermal treatments.

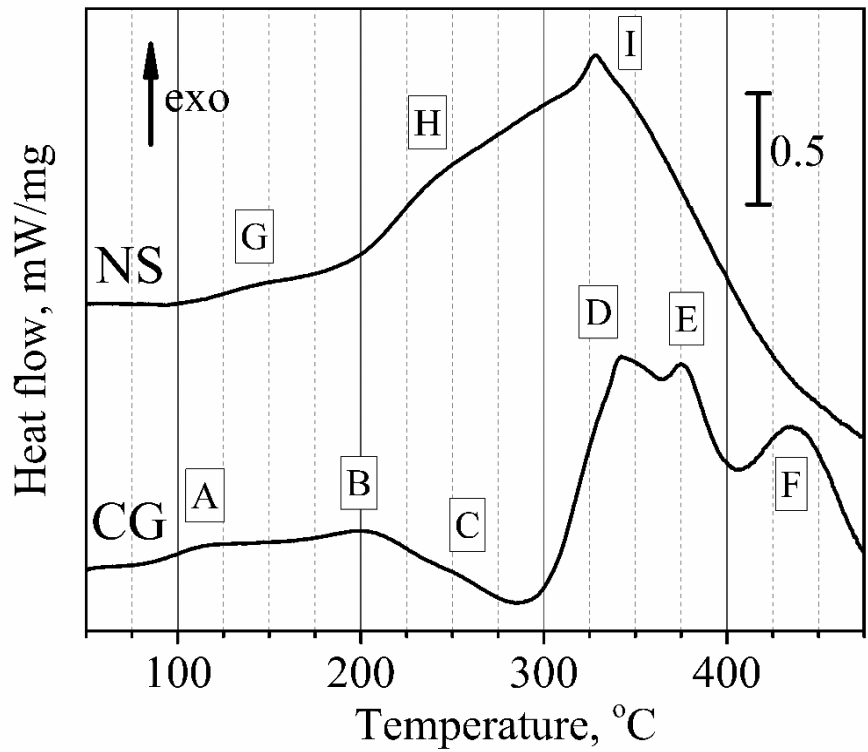

Fig. 8. DSC curves of CG and NS by HPT of Cu – 2Gl alloy. Peaks A – F are from the CG alloy and G – I are from the NS alloy and indicate the heat release peaks related to phase transformations observed during heating. The XRD patterns of CG and NS Cu-2Be alloy after HPT and different thermal treatments.

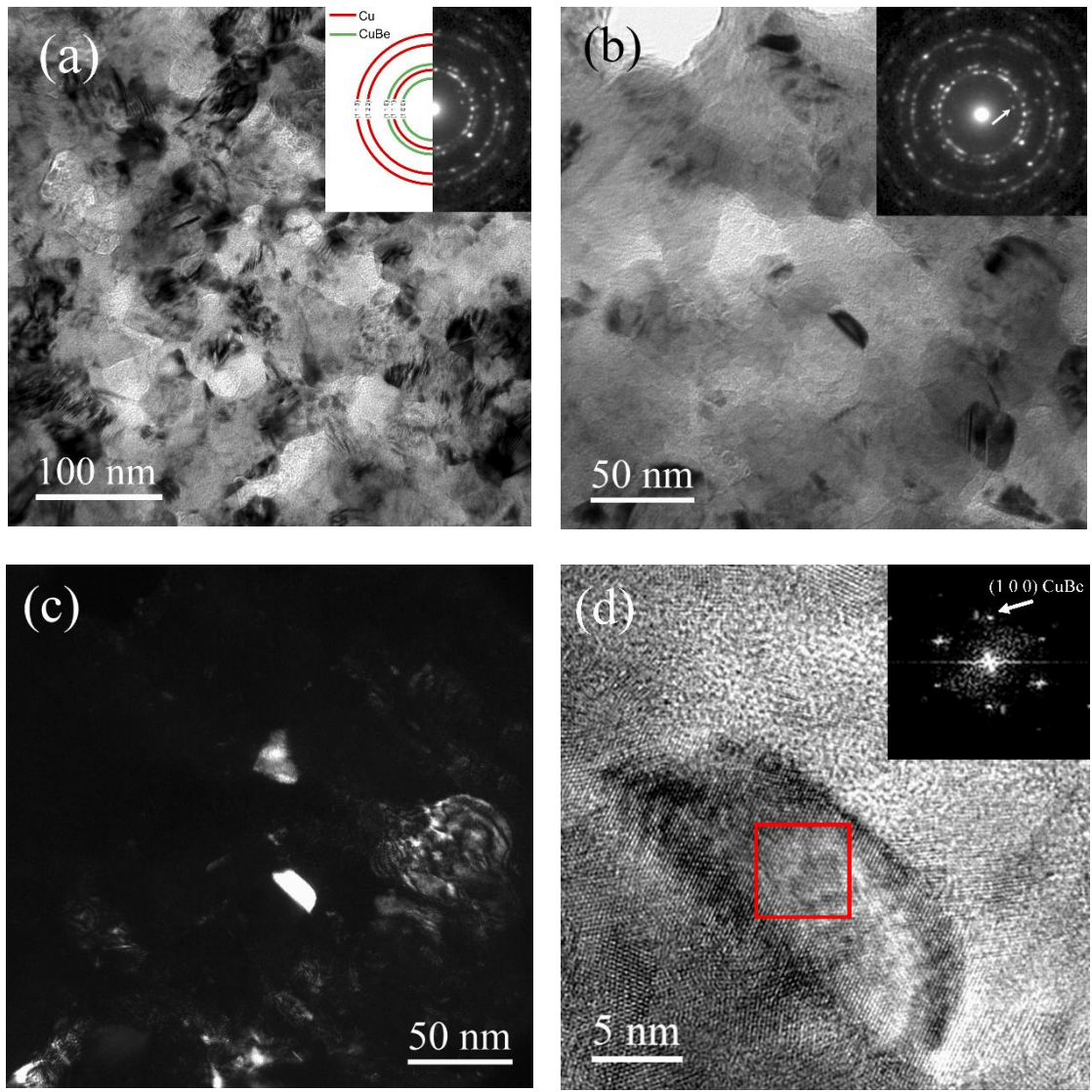

Fig. 9. (a) Microstructure of Cu-2Be alloy subjected to HPT and subsequent thermal treatment at 275°C for 30 minutes. Bright field image and diffraction pattern (b) with dark field image (c) of the same area obtained using selected CuBe (100) plain diffraction spot (arrow). (d) HRTEM image of CuBe particle with corresponding FFT of the area marked by red square

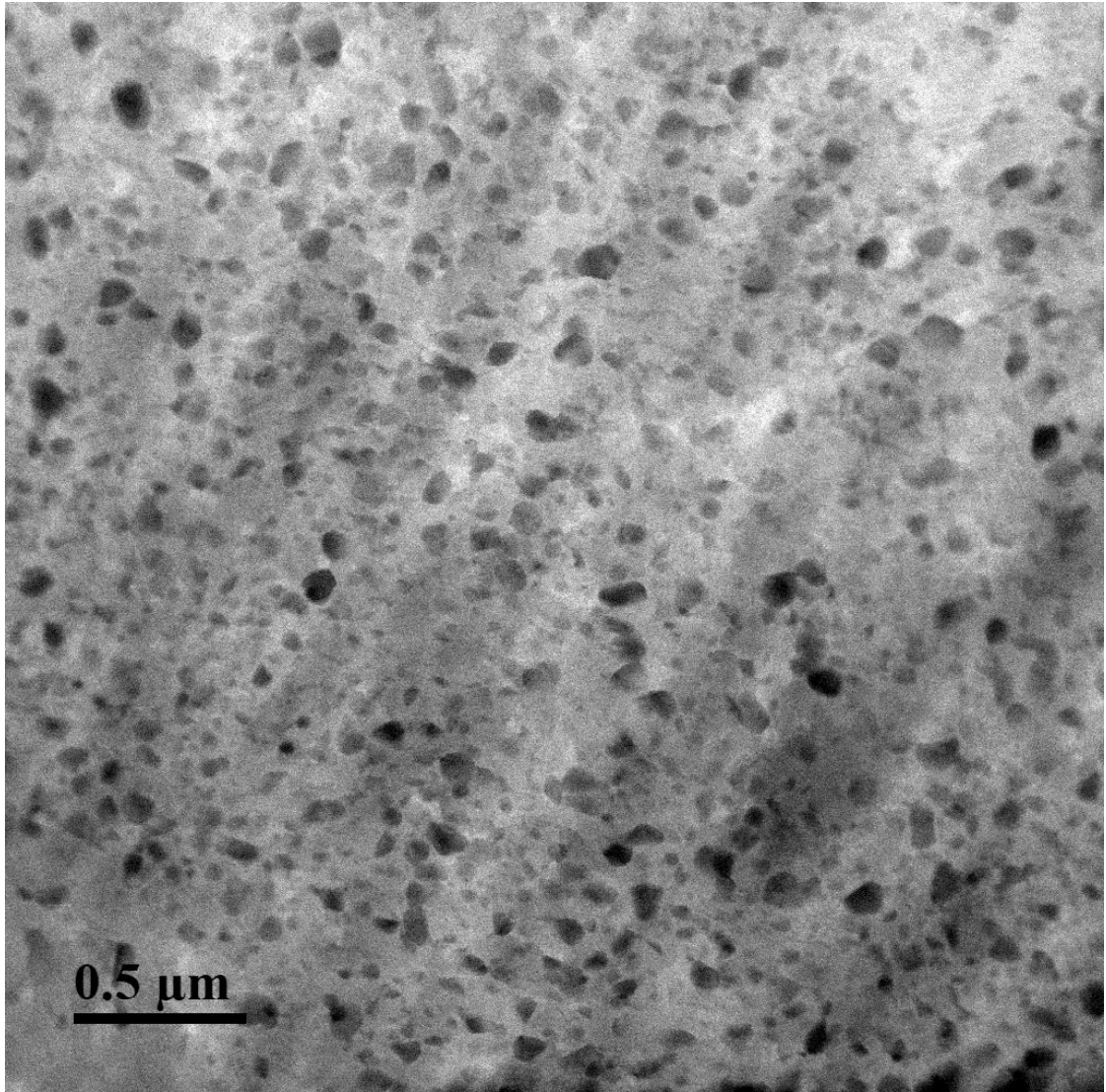
Fig. 10. A STEM HAADF image of NS Cu-2 wt.% Be after aging at 275°C for 10 hours.

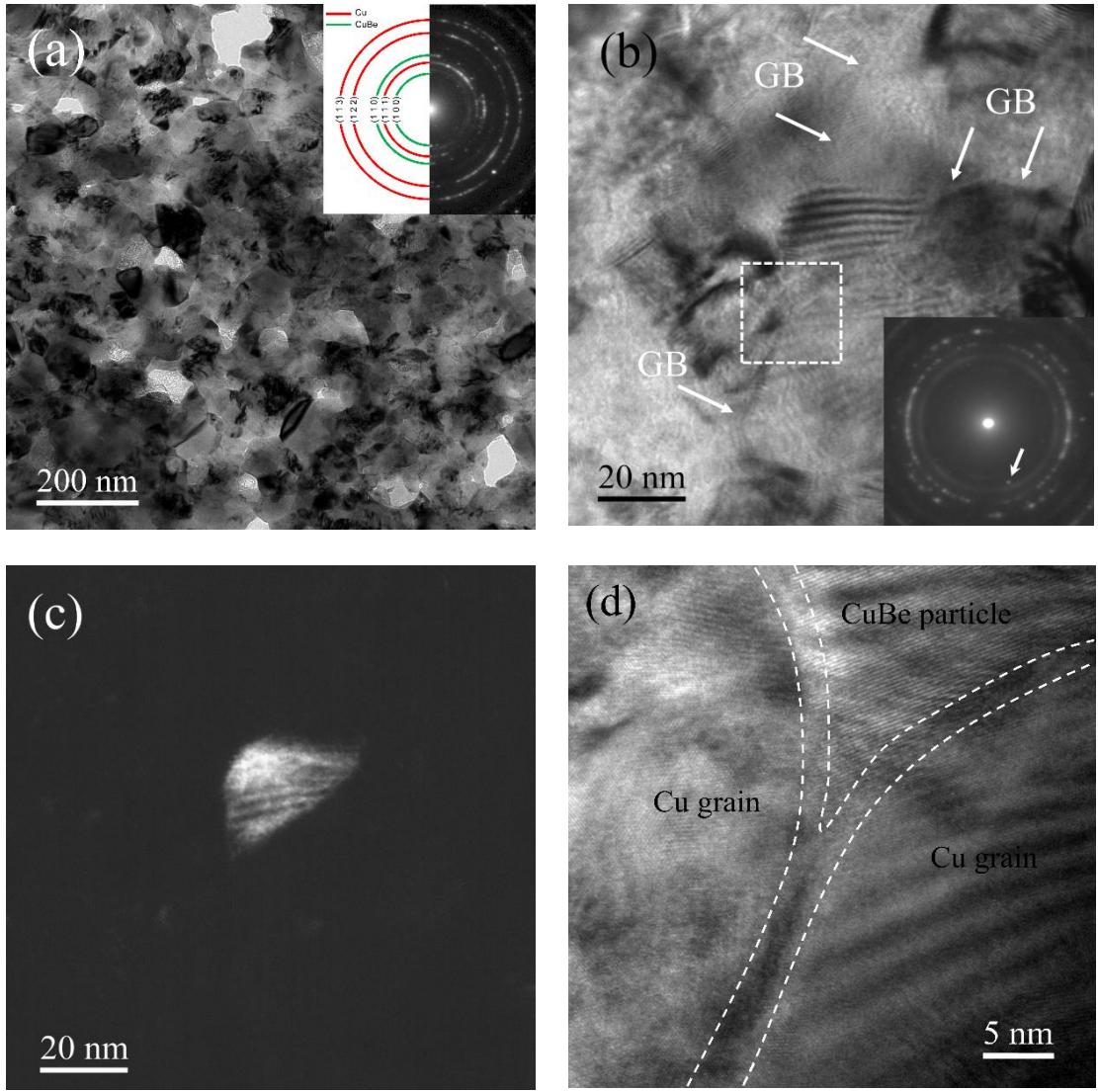

Fig. 11. (a) TEM image of the Cu-2Be alloy subjected to HPT and subsequent thermal treatment at 275ºC for 10 hours with related diffraction pattern. (b) Bright field image of the Cu-Be particle observed on the grain boundary. (c) Arrow on the diffraction pattern point on the (100) plane diffraction spot used to obtain dark field image. (d) HRTEM of the particle at the grain junction.